\documentclass[twocolumn,pra,amssymb,showpacs]{revtex4-1}
\usepackage{graphicx}
\usepackage{amsmath}
\usepackage{subfigure}

\newcommand{\be}{\begin{equation}}
\newcommand{\ee}{\end{equation}}
\newcommand{\ben}{\begin{eqnarray}}
\newcommand{\een}{\end{eqnarray}}

\usepackage[all]{xy}
\usepackage{amsfonts}
\usepackage{amssymb}
\usepackage{amscd}
\usepackage{amsthm}
\usepackage{latexsym}
\usepackage{amsbsy}

\newtheorem{theorem}{Theorem}[section]
\newtheorem{remark}[theorem]{Remark}

\newtheorem{lemma}[theorem]{Lemma}

\newcommand{\rnc}[2]{\renewcommand{#1}{#2}}
\rnc{\theequation}{\thesection.\arabic{equation}}
\rnc{\[}{\begin{equation}}
\rnc{\]}{\end{equation}}

\begin{document}

\title{Cyclic groups and quantum logic gates}
\author{Arash Pourkia$^1$, J. Batle$^2$ and C. H. Raymond Ooi$^3$}
\email{E-mail addresses: (AP) apourki@uwo.ca, (JB) jbv276@uib.es}
\affiliation{
$^1$ Mathematics division, College of Engineering, American University of the Middle East, 220 Dasman, 15453 Kuwait\\
$^2$Departament de F\'{\i}sica, Universitat de les Illes Balears, 07122 Palma de Mallorca, Balearic Islands, Europe\\
$^3$Department of Physics, University of Malaya, 50603 Kuala Lumpur, Malaysia\\\\}

\date{\today}

\begin{abstract}
We present a formula for an infinite number of universal quantum logic gates, which are $4$ by $4$ unitary solutions to
the Yang-Baxter (Y-B) equation. We obtain this family from a certain representation of the cyclic group of order $n$.
We then show that this {\it discrete} family, parametrized by integers $n$, is in fact, a small sub-class of a larger
{\it continuous} family, parametrized by real numbers $\theta$, of universal quantum gates.
We discuss the corresponding Yang-Baxterization and related symmetries in the concomitant Hamiltonian.
\end{abstract}


\maketitle

\section{Introduction}

Quantum correlations lie at the heart of quantum information theory and quantum computation. They are responsible for some tasks that possess no classical counterpart. Among those correlations,
entanglement is perhaps one of the most fundamental and non-classical feature exhibited by quantum systems \cite{LPS98,Galindo,NC00,LPS98,WC97,W98,E91,BBCJPW93,BW93,EJ96,BDMT98}.\newline

In recent years, a different approach to quantum entanglement has been developed, with the ultimate goal of achieving fault-tolerant quantum computation. In particular, the works of Kauffman and Lomonaco \cite{kaufman lomonaco 02, kaufman lomonaco 03, kaufman lomonaco 04}, on the connections between quantum entanglement, topological entanglement, and quantum computing, have brought the unitary solutions to the Yang-Baxter (Y-B) equation, i.e. equation \eqref{byb} below, to the center of attention.\newline

Let $V$ be a $n$ dimensional Vector (Hilbert) space over a field $F$ (for us $F=\mathbb{C}$ the field of complex numbers), and let $R:V\otimes V \rightarrow V\otimes V$ be a linear map. When $R$  is unitary (i.e.  $R^{-1}=R^{\dagger}$ the conjugate transpose of $R$), it could be considered as a quantum logic gate, in quantum computing. In the study of quantum entanglement in quantum computing, it is critically important when $R$ is entangling i.e. when creates entangled states from non-entangled ones \cite{kaufman lomonaco 02, kaufman lomonaco 03, kaufman lomonaco 04}.\newline

A linear map $R:V \otimes V \rightarrow V\otimes V$ is said to be a solution to  parameter-independent Y-B equation, if it satisfies the relation, \cite{kasslbook, majidbook}.
\[ (R\otimes I)(I\otimes R)(R\otimes I)=(I\otimes R)(R\otimes I)(I\otimes R), \label{byb}\]
where $I$ is the identity map on $V$. Since $R$ is a liner map one can present it as a $n^2$ by $n^2$ matrix for some basis of $V$. If $R$ is invertible, it provides an infinite family of braid group representations \cite{kasslbook, majidbook}. Which, in  turn, yields some invariants of links \cite{jones1985, kauffman1987, turaev 1992}. This, in a sense, is where the topological entanglement being studied \cite{kaufman lomonaco 02, kaufman lomonaco 03, kaufman lomonaco 04}.\newline

The so called unitary braiding operators, i.e. unitary solutions to Y-B equations, and the relations between quantum and topological entanglement in quantum computing, have been studied extensively by many authors, in the last decate or so. For example some of the works that we will refer to in this paper are in \cite{kaufman etal 05-1, kaufman etal 05-2, brylinskis 02, jchen et al 07, dye 03, pinto et al 13, alagic et al 15, lho et al 10}. \newline

On the other hand, it is a very well known fact, \cite{kasslbook, majidbook} that a natural source of solutions to
 Y-B equation is from quasitriangular Hopf algebras. In a quasitriangular Hopf algebra, $(H, R)$, $H$ is a Hopf algebra and $R=\sum_{i} \mathfrak{r}_i\otimes \mathfrak{s}_i$ (usually summation understood and eliminated) is an invertible element in $H \otimes H$ satisfying certain relations \cite{kasslbook, majidbook}. This element $R$ satisfies the following version of parameter-independent Y-B equation,
 \[R_{12}R_{13}R_{23}=R_{23}R_{13}R_{12}, \label{paramindepybealg}\]
 where $R_{12}=\sum_{i} \mathfrak{r}_i\otimes \mathfrak{s}_i\otimes 1$, $R_{13}=\sum_{i} \mathfrak{r}_i\otimes 1\otimes \mathfrak{s}_i$, and $R_{23}=\sum_{i} 1\otimes \mathfrak{r}_i\otimes \mathfrak{s}_i$.

 This property of $R$ implies that $\tau R$ gives rise to representations of {\it Artin braid group} $B_n$ or correspondingly to a solution to Y-B equation \eqref{byb}. Here $\tau$ is the flip (swap) map given by, $\tau(x\otimes y)=y\otimes x$. \newline

In this paper we focus on two dimensional vector spaces $V$ over, $\mathbb{C}$, the field of complex numbers. In this case, $R:V\otimes V \rightarrow V\otimes V$ can be represented by a $4$ by $4$ matrix with entries in $\mathbb{C}$, with respect to a basis of $V$. \newline

Usually, the preferred basis for two qubit states or gates is the so called computational basis
$\{ |00\rangle ,|01\rangle,|10\rangle,|11\rangle \}$.
In our case, it will prove at some point convenient to employ the so called Bell basis of
maximally correlated states, which are of the form

\begin{equation} \label{bellbasis}
|\Phi^{\pm} \rangle= \frac{(|00\rangle \pm |11\rangle)}{\sqrt{2}}, |\Psi^{\pm} \rangle= \frac{(|01\rangle \pm |10\rangle)}{\sqrt{2}}.
\end{equation}

\noindent The employment of a unitary solutions to the Y-B equation will eventually bring us to a description of Hamiltonians in terms of the computational basis, which shall be represented in the form of tensor products of the generators of the $su(2)$-group, that is, the familiar Pauli matrices.\newline

This paper is organized as follows. In Section \ref{Bn}, we obtain quantum logic gates from cyclic groups, $C_n$, for any order $n$, via the quasitriangular structure on their group Hopf algebra . The corresponding proof is given in the Appendix. The corresponding Yang-Baxterization is performed in Section \ref{ybxization}. The analysis of the ensuing family of Hamiltonians and the concomitant physical applications appears is Section \ref{hamilton}. General continuous quantum gates are studied in Section \ref{genral}. A second approach to quantum logic gates from cyclic groups is given in Section \ref{Cnsecondapproach}. Finally, some conclusions are drawn in Section \ref{Conclusions}.

\section{Quantum gates correspond to cyclic groups} \label{Bn}

In this section we state one of our main lemmas, whose outcome help us to obtain an infinite family of $4$ by $4$ unitary solutions to Yang-Baxter (Y-B) equation \eqref{byb}, from the cyclic group of order $n$. The matrices $\mathfrak{B}_n$ in this family are entangling universal logic gates, for $n\neq 2, 4$. We then prove that a deformation of $\mathfrak{B}_n$ by a phase factor has all the above mentioned properties as well.

We recall  \cite{kasslbook, majidbook}, a natural source of solutions to the (parameter-independent) Y-B equation \eqref{byb}, and correspondingly a source for representations of Artin braid group $B_n$ (don't be confused with  $\mathfrak{B}_n$ in the present paper ), is from quasitriangular Hopf algebras $(H, R)$. Where, $R=\sum_{i} \mathfrak{r}_i\otimes \mathfrak{s}_i$ is an invertible element in $H \otimes H$ satisfying certain relations. It turns out $R$ satisfies the, sometimes called algebraic (parameter-independent) Y-B equation \eqref{paramindepybealg}. This in turn implies that, $\tau R$ is a solution to, sometimes called braided (parameter-independent) Y-B equation \eqref{byb}, where $\tau$ is the flip (swap) map given by, $\tau(x\otimes y)=y\otimes x$. \newline . \newline

\begin{lemma}\label{Bn first approach}
Let $C_n=\{1, s, s^2,\cdots ,s^{n-1}\}$ be the cyclic group of order $n$, for any $n\geq 2$, with the generator $s$, satisfying the relation $s^n=s^0=1$. Let $\omega=e^{\frac{2\pi i}{n}}$, and,
\begin{equation}
 R= \frac{1}{n}\sum_{a,b=0}^{n-1} \omega^{-ab}s^a \otimes s^b
\end{equation}
be the, well known \cite{majidbook}, non-trivial quasitriangular structure on the group Hopf algebra $\mathbb{C}C_n$. If we use the following $2$ by $2$ matrix representation of elements of $C_n$,
\[s^a=\begin{pmatrix}
\cos {\frac{2a\pi}{n}}& -\sin {\frac{2a\pi}{n}}\\
\sin {\frac{2a\pi}{n}}& \cos {\frac{2a\pi}{n}}
\end{pmatrix}, \qquad a=0,1,\cdots , n-1 \]
then we will obtain a $4$ by $4$ representation of $R$ that has the following form,
\begin{align}
R=
\begin{pmatrix}
\cos {\frac{2\pi}{n}}&0&0&-i\sin {\frac{2\pi}{n}} \\
0&\cos {\frac{2\pi}{n}}&i\sin {\frac{2\pi}{n}}&0 \\
0&i\sin {\frac{2\pi}{n}}&\cos {\frac{2\pi}{n}}&0 \\
-i\sin {\frac{2\pi}{n}}&0&0&\cos {\frac{2\pi}{n}}
\end{pmatrix}
\end{align}\\
\end{lemma}

The proof of the above Lemma is very long and computationally involved. We present the full proof in the Appendix.\newline

Next we apply the SWAP gate $S=\begin{pmatrix}
1& 0&0& 0\\
0& 0&1& 0\\
0& 1&0& 0\\
0& 0&0& 1
\end{pmatrix}$ to $R$ to obtain,

\begin{align}\nonumber
&\mathfrak{B}_n=S\circ R=\\
&\begin{pmatrix}
\cos {\frac{2\pi}{n}}&0&0&-i\sin {\frac{2\pi}{n}} \\
0&i\sin {\frac{2\pi}{n}}&\cos {\frac{2\pi}{n}}&0 \\
0&\cos {\frac{2\pi}{n}}&i\sin {\frac{2\pi}{n}}&0 \\
-i\sin {\frac{2\pi}{n}}&0&0&\cos {\frac{2\pi}{n}}
\end{pmatrix}
\end{align}

In other words, $\mathfrak{B}_n$ has the following form,
\begin{align} \label{generelformofBn}
\mathfrak{B}_n=\begin{pmatrix}
\alpha&0&0&\beta \\
0&-\beta&\alpha&0 \\
0&\alpha&-\beta&0 \\
\beta&0&0&\alpha
\end{pmatrix}
\end{align}
with, $\alpha=\cos {\frac{2\pi}{n}}$ and $\beta= -i\sin {\frac{2\pi}{n}}$.\newline

It is clear that $\mathfrak{B}_n$ is unitary. It could be understood that $\mathfrak{B}_n$  is a solution to Y-B equation because it is obtained from a quasitriangular structure. However, in Section \ref{genral} we see that being a solution to Y-B equation (and the unitarity) could be obtained, by direct calculation, for more general family, of which $\mathfrak{B}_n$ is only a small subfamily.

\begin{remark}
The fact that $\mathfrak{B}_{n}$ is obtained from the cyclic group of order $n$, is not obvious at all. That is the subject of Lemma \ref{Bn first approach}. \newline
\end{remark}

Except for $n=2,4$, any $\mathfrak{B}_n$, is entangling \cite{kaufman lomonaco 04}. Because, for example,
\[\mathfrak{B}_n (|00\rangle)
= \cos {\frac{2\pi}{n}}|00\rangle
 -i\sin {\frac{2\pi}{n}} |11\rangle
\nonumber\]
Thus, by Brylinskis' Theorem \cite{brylinskis 02}, $\mathfrak{B}_n$ is a universal quantum gate, for $n\neq 2,4$. \newline

From the general form for $\mathfrak{B}_n$, for any $n$, one notices that the $\mathfrak{B}_n$ gate can be decomposed as $\alpha S + \beta S'$, with

\begin{align}
S=
\begin{pmatrix}
1&0&0&0 \\
0&0&1&0 \\
0&1&0&0 \\
0&0&0&1
\end{pmatrix}
\end{align}

\noindent and

 \begin{align}
S'=
\begin{pmatrix}
0&0&0&1 \\
0&-1&0&0 \\
0&0&-1&0 \\
1&0&0&0
\end{pmatrix}
\end{align}

\noindent As known, the problem of finding solutions to the Yang-Baxter equation that are
unitary turns out to be surprisingly difficult. The classification of all such matrices of size $4 \times 4$, in done in \cite{dye 03}, based on the work in \cite{hietarinta}. We see that $S$ and $S'$  belong to two types of those classified ones. The fact that $\mathfrak{B}_n$ as a linear superposition of (unitary) solutions to the Yang-Baxter equation is also a (unitary) solution, is interesting on its own. Because this is not true in general.

In terms of Bell-basis projectors, $S=I_{4 \times 4} - 2|\Psi^{-}\rangle \langle \Psi^{-}|$ and $S'= 2|\Phi^{+}\rangle \langle \Phi^{+}| - I_{4 \times 4}$.
The entire $\mathfrak{B}_n$ gate thus reads $(\alpha-\beta) I_{4 \times 4} - 2\alpha|\Psi^{-}\rangle \langle \Psi^{-}| + 2\beta|\Phi^{+}\rangle \langle \Phi^{+}|$. It is clear
from this form that the entangling gate $\mathfrak{B}_n$ is linear in the Bell basis.

Soon, we shall discuss the corresponding Yang-Baxterization $\breve{R_n}(x)$ \cite{ychen etal 1991} of $\mathfrak{B}_n$, and related symmetries in the concomitant Hamiltonian in Sections \ref{ybxization} and \ref{hamilton}. After Yang-Baxterization for $\mathfrak{B}_n$, we shall notice that only coefficients $\alpha$ and $\beta$ will change as a function of $x$. This fact implies that the study of how gate $\mathfrak{B}_n$ can be implemented in terms of
spin-$\frac{1}{2}$ operators (the Pauli matrices) reduces to $S$ and $S'$. Where, $S$ is exactly
$P_{i,i+1} = \frac{1}{2}(1+{\vec \sigma}_i\cdot {\vec \sigma}_{i+1})$, the permutation operator for spins $i$ and $i+1$.
The operator obviously satisfies $P_{i,i+1}^2 = 1$ and is a solution to the following braid relations:

\begin{eqnarray}
  b_{i, i+1}  b_{i+1, i+2} b_{i, i+1} & = & b_{i+1, i+2} b_{i, i+1} b_{i+1, i+2},
\; i \leq N-2, \nonumber\\
  b_{i, i+1} b_{j, j+1} & = & b_{j, j+1} b_{i, i+1}, \;\;
   | i - j | \ge 2.
\end{eqnarray}

\noindent Further insight into the whole operator $\mathfrak{B}_n$, will be obtained when it is written down as,

\begin{equation}
\mathfrak{B}_n\,=\,(\alpha-\beta) \frac{1}{2}(1+{\vec \sigma}_a \cdot {\vec \sigma}_{b}) \,+\,\beta [\sigma_{x}^a \otimes \sigma_{x}^b +
\sigma_{z}^a \otimes \sigma_{z}^b ].
\end{equation}

\subsection{$q$-deformation of $\mathfrak{B}_n$}
We show that a deformation of $\mathfrak{B}_n$ by a phase factor $q=e^{i\varphi}$, is still a unitary solution to Y-B equation. It will be entangling, and as a result a universal quantum gate \cite{brylinskis 02}.

\begin{lemma}
If for simplicity we write $\mathfrak{B}_n$ as:
\[\mathfrak{B}_n=
\begin{pmatrix}
a&0&0&-ib \\
0&ib&a&0 \\
0&a&ib&0 \\
-ib&0&0&a
\end{pmatrix} \label{deformedBnab}\] where, $a=\cos {\frac{2\pi}{n}}$, $b=\sin {\frac{2\pi}{n}}$, then the following deformation of $\mathfrak{B}_n$ is also unitary and a solution to Y-B equation \eqref{byb}.
\[\mathfrak{B}_{n, \varphi}:=\begin{pmatrix}
a&0&0&-ibe^{i\varphi} \\
0&ib&a&0 \\
0&a&ib&0 \\
-ibe^{-i\varphi}&0&0&a
\end{pmatrix} \label{deformedBn}\]
\end{lemma}
\proof
Let
$Q=\begin{pmatrix}
e^{i\frac{\varphi}{2}}&0 \\
0&1
\end{pmatrix}$.
It is clear that $Q$ is unitary, and it is easy to see $(Q\otimes Q)\mathfrak{B}_n(Q\otimes Q)^{-1}=\mathfrak{B}_{n, \varphi}$. Then we will have the desired result, because of the following well known fact  \cite{kasslbook, majidbook}. If $B$ is a solution to Y-B equation, and if $Q: V \rightarrow  V$ is an invertible map, then $(Q\otimes Q)B(Q\otimes Q)^{-1}$ is also a solution to Y-B equation .\newline
As expected, $\mathfrak{B}_n$ is an especial case of $\mathfrak{B}_{n, \varphi}$, when $\varphi=0$.

\section{Yang-Baxterization}\label{ybxization}
In this section, similar to the methods used in \cite{kaufman etal 05-1, jchen et al 07}, we wish to obtain the Yang-Baxterization \cite{ychen etal 1991} correspond to $\mathfrak{B}_{n, \varphi}$. This is the process of going from a parameter-independent solution to Y-B equation $\mathfrak{B}_{n, \varphi}$,  to a parameter-dependent solution, $\breve{R}(x)$, where $x$ is called the (spectral) parameter \cite{ychen etal 1991}. This will allow us to discuss the Hamiltonian and other  physical aspects related to $\mathfrak{B}_{n, \varphi}$, in Section \ref{hamilton}. For that, we will eventually transform $\breve{R}(x)$ into $\breve{R}(\theta,\varphi)$ in \eqref{ybterization of Bn}. However, in Section \ref{genral}, we will see that $\breve{R}(\theta,\varphi)$ can also be obtained directly, without going through the Yang-Baxterization process. \newline

Recall from the last section,
\[\mathfrak{B}_{n, \varphi}=\begin{pmatrix}
a&0&0&-ibe^{i\varphi} \\
0&ib&a&0 \\
0&a&ib&0 \\
-ibe^{-i\varphi}&0&0&a
\end{pmatrix}, \] where, $a=\cos {\frac{2\pi}{n}}$, $b=\sin {\frac{2\pi}{n}}$.\newline

After very simple algebra, we find three distinct non-vanishing eigenvalues for $\mathfrak{B}_n$, namely $a+bi$, $-a+bi$, and $a-bi$. Thus, $\mathfrak{B}_{n, \varphi}$ fits into the case(s) with three distinct non-vanishing eigenvalues $\lambda_1, \lambda_2, \lambda_3$. The formula for Yang-Baxterization for such a $B$ is given by \cite{ychen etal 1991}:
{\small \begin{equation}\label{yb-ization3lamda}
\breve{R}(x)= \lambda_1\lambda_3 x(x-1)B^{-1} + (\lambda_1+\lambda_2+\lambda_3+\lambda_1\lambda_3\lambda^{-1}_2)x I-(x-1)B
\end{equation}}
There are six different possible ways to assign $\lambda_1, \lambda_2, \lambda_3$ but usually one obtains (at most) three different type of $\breve{R}(x)$-matrices. \newline

For the current purpose, let us assign $\lambda_1=a+bi,\, \lambda_2=-a+bi,\, \lambda_3=a-bi$. Then, $\lambda_1\lambda_3=1$, $\lambda_1+\lambda_2+\lambda_3=\lambda_1=-(\lambda_2)^{-1}$. Applying these into formula \eqref{yb-ization3lamda}, we have:
\[\breve{R}(x)= x(x-1)\mathfrak{B}_{n, \varphi}^{-1}-(x-1)\mathfrak{B}_{n, \varphi}\]\newline

Using $\mathfrak{B}_{n, \varphi}$ as above and $\mathfrak{B}_{n, \varphi}^{-1}=\mathfrak{B}_{n, \varphi}^\dagger$ (the conjugate transpose of $\mathfrak{B}_{n, \varphi}$), we get,
\begin{align}\label{R(x)}
 &\breve{R}(x)=\nonumber\\
& {\small\begin{pmatrix}
a(x-1)^2& 0&0& bi(x^2-1)e^{i\varphi}\\
0& -bi(x^2-1)&a(x-1)^2& 0\\
0& a(x-1)^2&-bi(x^2-1)& 0\\
bi(x^2-1)e^{-i\varphi}& 0&0& a(x-1)^2
\end{pmatrix}}
\end{align}
 Notice that, $\breve{R}(0)=\mathfrak{B}_n$ as expected in the process of Yang-Baxterization, and  $\breve{R}(1)=0$ which is un-interesting.\newline

Next we Impose the unitarity condition, $\breve{R}(x)\breve{R}(x)^\dagger=\breve{R}(x)^\dagger\breve{R}(x)=\rho I$, where $\rho$ is a normalization factor for the $\breve{R}(x)$-matrix , and where
{\footnotesize
\arraycolsep=3pt
\medmuskip = 1mu
\[\breve{R}(x)^\dagger=\begin{pmatrix}
a(\bar{x}-1)^2& 0&0& -bi(\bar{x}^2-1)e^{i\varphi}\\
0& bi(\bar{x}^2-1)&a(\bar{x}-1)^2& 0\\
0& a(\bar{x}-1)^2&bi(\bar{x}^2-1)& 0\\
-bi(\bar{x}^2-1)e^{-i\varphi}& 0&0& a(\bar{x}-1)^2
\end{pmatrix} \]}\newline

Imposing the unitarity condition, forces the following restrictions on $x$ and $\rho$:
\[a^2(x-1)^2 (\bar{x}-1)^2 +b^2(x^2-1)(\bar{x}^2-1)=\rho\]
\[-abi(x-1)^2(\bar{x}^2-1)+abi(x^2-1)(\bar{x}-1)^2=0\]
From these two equations, we find that $x$ must be a real number, $x\neq 1$, and
\[\rho=a^2(x-1)^4+b^2(x^2-1)^2 \label{rho}\]

\begin{remark}
The arrangement of eigenvalues we used in above is, in a sense, equivalent to the case with two distinct eigenvalues.  We can also assign, $\lambda_1, \lambda_2, \lambda_3$ either as  $\lambda_1=-a+bi,\, \lambda_2=a-bi, \, \lambda_3=a+bi$, for which result will be similar to what we have shown above. Or as $\lambda_1=-a+bi,\, \lambda_2=a+bi, \, \lambda_3=a-bi$, which is the case with really three distinct eigenvalues. We shall explore this case elsewhere.
\end{remark}


\section{Hamiltoninan}\label{hamilton}
To proceed with Hamiltonian, the normalized form $\breve{R}(x)$ \eqref{R(x)}, i.e. $\frac{1}{\sqrt{\rho}}\breve{R}(x)$ must be used,
where $\rho$ is given by the formula \eqref{rho}. Here is the simplified version of $\frac{1}{\sqrt{\rho}}\breve{R}(x)$ ($x\neq 1$).

\begin{align}\label{nomalRx}\nonumber
&\frac{1}{\sqrt{\rho}}\breve{R}(x)=\\
&{\footnotesize
\arraycolsep=2pt
\medmuskip = 1mu
k\begin{pmatrix}
a(x-1)& 0&0& bi(x+1)e^{i\varphi}\\
0& -bi(x+1)&a(x-1)& 0\\
0& a(x-1)&-bi(x+1)& 0\\
bi(x+1)e^{-i\varphi}& 0&0& a(x-1)
\end{pmatrix}}
\end{align}
where $k=\frac{1}{\sqrt{a^2(x-1)^2+b^2(x+1)^2}}$\newline

Let us further analyze $\frac{1}{\sqrt{\rho}}\breve{R}(x)$. Introducing a new angle $\theta$, which will be dependent on $n$ and $x$, we define
$\cos\theta=a(x-1)/\sqrt{a^2(x-1)^2+b^2(x+1)^2}$ and $\sin\theta=b(x+1)/\sqrt{a^2(x-1)^2+b^2(x+1)^2}$. The assignment of $\cos\theta$ and $\sin\theta$ is arbitrary and could
be done the other order, with no difference in the following results. The ensuing gate from (\ref{nomalRx}) will read as

\[\breve{R}_{u,v}(\theta,\varphi)=\begin{pmatrix}
\cos\theta&0&0&i\sin\theta e^{i\varphi} \\
0&-i\sin\theta&\cos\theta&0 \\
0&\cos\theta&-i\sin\theta&0 \\
i\sin\theta e^{-i\varphi}&0&0&\cos\theta
\end{pmatrix} \label{ybterization of Bn}\]

\noindent where from now on $u,v$ will refer to qubit $u$ and qubit $v$. Gate $\breve{R}_{u,v}(\theta,\varphi)$, when acting upon any element of the computational basis $\{|00\rangle,|01\rangle,|10\rangle,|11\rangle\}$, the outcome are states whose concurrence (a measure of bipartite entanglement) is always $C=|\sin2\theta|$, regardless of $\varphi$. When $\theta=\frac{\pi}{4}$ we recover the maximally correlated Bell basis. On the contrary, when $\theta=\frac{\pi}{2}$ the gate is unentangling, which happens only for the cyclic group order $n=2,4$ ($x=1$ is not relevant).

With two parameters $\theta$, $\varphi$ in $\breve{R}_{u,v}(\theta,\varphi)$, we can choose which one to play the role of time. 

Let us assume now $\varphi$ to be time-dependent (usually in a linear fashion) while $\theta$ be time-independent.
We thus obtain a Hamiltonian through the
unitary transformation $\breve{R}_{u,v}(\theta,\varphi)$ as
\begin{eqnarray}\label{hamiltonian}
&&H(\theta,\varphi)=i\hbar\frac{\partial
\check{R}_{u,v}(\theta,\varphi)}{\partial
t}\check{R}^\dag_{u,v}(\theta,\varphi).
\end{eqnarray}

\noindent From now on we assume $\hbar=1$. The Hamiltonian reads

$H(\theta,\varphi)={\dot \varphi} \sin\theta \begin{pmatrix}
-\sin\theta&0&0&-i \cos\theta e^{i\varphi} \\
0&0&0&0 \\
0&0&0&0 \\
i\cos\theta e^{-i\varphi}&0&0&\sin\theta
\end{pmatrix}$,

\noindent which is equivalent, using spin-1/2 ladder operators $S^\pm=S^x\pm i S^y$, to

\begin{align} \label{hamiltonianSpin}
H(\theta,\varphi)&= - {\dot \varphi} \sin\theta [
\sin\theta (S^z_u \otimes {\bf 1}_{v}+ {\bf 1}_{u} \otimes
S^z_{v}) \\\nonumber
& \qquad+  i\cos\theta (e^{i\varphi} S^+_u \otimes S^+_{v}- e^{-i\varphi} S^-_u
\otimes S^-_{v}) ].
\end{align}

\noindent Notice that, as far as the physical meaning of the ensuing Hamiltonian is concerned, there is no substantial difference in
switching the roles of $\sin\theta$ and $\cos\theta$, or in flipping the sign of the phase $\varphi$.

Now, if we recall the definition of the concurrence $C=|\sin2\theta|$ once $\breve{R}_{u,v}(\theta,\varphi)$ is applied to the standard
computational basis, we can rewrite (\ref{hamiltonianSpin}) as

\begin{eqnarray}\label{Hfinal}\nonumber
H(\theta,\varphi)&=&H_{free}\,+\, H_{int}\\
&=& - {\dot \varphi} \sin^2\theta (S^z_u \otimes {\bf 1}_{v}+ {\bf 1}_{u} \otimes S^z_v)\\\nonumber
&& \mp i \dot{\frac{\varphi}{2}} C (e^{i\varphi} S^+_u \otimes S^+_{v}- e^{-i\varphi} S^-_u
\otimes S^-_{v}),
\end{eqnarray}

\noindent with the $\mp$ depending on the sign of $\sin2\theta$. The previous form for the Hamiltonian unveils the corresponding
entangling nature: the cyclic group leads, after a $q-$deformation of $\mathfrak{B}_n$, to a Hamiltonian which is split into
two contributions: i) the first one $H_{free}$, where particles do not interact (this does not imply that they may not have entanglement), and
ii) the second one $H_{int}$, an interacting contribution. The interacting part of the Hamiltonian is always present (except for $n=2,4$)
whenever $\breve{R}_{u,v}(\theta,\varphi)$ is entangling any pure product state ($C \neq 0$). Thus, in this case, physical interaction between
qubits automatically means entanglement.

It is noteworthy to note that, Hamiltonian (\ref{Hfinal}) looks very similar to a recent proposal for the realization of an exact CNOT gates with a single nonlocal rotations
for quantum-dot qubits \cite{Pal}. In our case $ {\dot \varphi} \sin^2\theta $ would play the role of an
effective external magnetic field $B$. The interacting part, linear combinations of tensors products of $\sigma_x$ and $\sigma_y$, may represent
capacitive coupling between qubits expressed in terms of a Coulomb interaction between electric charges on pairs of dots belonging to different qubits.
The corresponding coupling constants would be either ${\dot \varphi}C\cos\varphi$ or ${\dot \varphi}C\sin\varphi$.\newline

If $\theta$ is time-dependent while $\varphi$ is time-independent, one can
construct a Hamiltonian which is antidiagonal in the computational basis, reading as
antidiag$(-{\dot \theta} e^{-i\varphi},{\dot \theta} ,{\dot \theta} ,-{\dot \theta} e^{i\varphi})$.
This Hamiltonian has the interesting property that its eigenvectors are the usual Bell states. When applied to $|\Psi^+\rangle$ and to
$|\Phi^+\rangle$, the corresponding outcomes are the same states, with eigenenergy ${\dot \theta}$. When applied to $|\Psi^-\rangle$ and to
$|\Phi^-\rangle$, the eigenenergies are $-{\dot \theta}$. This particular Hamiltonian converts Bell-energy eigenstates,
which are maximally correlated, into a subset of them.

\noindent The $\theta$ time-dependent (actually ${\dot \theta}$-dependent) Hamiltonian is expressed in terms of the Pauli matrices as:

\begin{align}\label{Hfinaltheta}
H({\dot \theta} , \varphi) / {\dot \theta}  &= \frac{1}{2} (\sigma_x\otimes\sigma_x + \sigma_y\otimes\sigma_y)\\ \nonumber
&-(e^{i\varphi} S^+_u \otimes S^+_{v}+e^{-i\varphi} S^-_u \otimes S^-_{v}).
\end{align}

\noindent Notice the similarity with (\ref{Hfinal}) when $\sin \theta=0$ and $C=1$. However, and regardless of the eigenstates of (\ref{Hfinaltheta})
being maximally entangled, (\ref{Hfinaltheta}) does not generate entangled states from disentangled
ones. Summing up, starting from $\breve{R}_{u,v}(\theta,\varphi)$, the time dependency either on $ \varphi$ or $\theta$
defines two concomitant Hamiltonians whose entangling properties are very different.\newline

Remarkably, our Hamiltonian $H(\theta,\varphi)$ is intimately related to the one obtained
in Ref. \cite{jchen et al 07}.
If we let the transformations $\{\theta \rightarrow \theta+\pi/2,\varphi \rightarrow -\varphi + \pi/2\}$ act
on $H(\theta,\varphi)$, we do exactly obtain the one discussed in \cite{jchen et al 07}.
However, \cite{jchen et al 07} discusses the $q$-deformation of the change of basis matrix $U$ from
the computational basis $\{|00\rangle,|01\rangle,|10\rangle,|11\rangle\}$ to the
Bell basis $\{|\Phi^{+}\rangle,|\Phi^{-}\rangle,|\Psi^{+}\rangle,|\Psi^{-}\rangle\}$. $q-$deformed $U$ is then
understood as a braiding operator, first discovered by Kauffman and Lomonaco \cite{kaufman lomonaco 02} during the investigation of the relationships among
quantum entanglement, topological entanglement and quantum computation by means of braiding
operators and Yang-Baxter equations.

We have to stress the fact that although the final Hamiltonians almost look the same, the way they are obtained are totally different.
In our case we use the cyclic groups and the corresponding $q$-$\mathfrak{B}_n$, which has nothing to do with the change of basis matrix $U$.

\subsection{Application: calculation of the Pancharatnam-Berry phase}

The Hamiltonian $H(\theta,\varphi)$, which is $\varphi$ time-dependent, has interesting echoes in the theory of geometric phases,
in particular the Berry phase \cite{Berry}.
The Berry phase is a phase difference acquired over the course of a cycle, when a system is subjected to cyclic adiabatic processes, which
results from the geometrical properties of the parameter space of a given Hamiltonian. The strong link with quantum information
theory comes with the so called {\it non-abelian} Berry phases. If implemented,
they may open entirely new possibilities for robust quantum information processing

Let us
find out first the eigenenergies and eigenstates of $H(\theta,\varphi)$. The eigenenergies are given by $\pm{\dot \varphi}\sin\theta$, with
eigenvectors (in the computational basis) $|\lambda^{\pm}(\theta,\varphi)\rangle\,=\,[i f^{\pm}(\theta)e^{i\varphi},0,0,g^{\pm}(\theta)]^{T}$ and
$f^{\pm}(\theta)=\frac{1}{\sqrt{2}}\sqrt{1\pm\sin\theta},\,g^{\pm}(\theta)=\frac{1}{\sqrt{2}} \frac{\cos\theta}{\sqrt{1\pm\sin\theta}}$. The states
corresponding to zero energy are not relevant here.

Now, if we calculate the concurrence for $|\lambda^{\pm}(\theta,\varphi)\rangle$, we easily obtain that it is equal to $C=|\cos\theta|$. According to the
theory of Berry phases, if we let $\varphi(t)$ evolve adiabatically from $0$ to $2\pi$, the corresponding Berry phase
for $|\lambda^{\pm}(\theta,\varphi)\rangle$ is given by

\begin{equation}
\gamma_{\pm}\,=\,i \int_0^T \langle \lambda^{\pm}(\theta,\varphi)| \,\frac{\partial}{\partial t} |\lambda^{\pm}(\theta,\varphi)\rangle\,dt\,=\,-\pi(1\pm\sin\theta).
\end{equation}

\noindent We can further relate the Berry phases $\gamma_{\pm}$ with the concurrence $C$ of the corresponding states
$|\lambda^{\pm}(\theta,\varphi)\rangle$ in the following fashion

\begin{equation} \label{BerryC}
\gamma_{\pm}\,=\,-\pi(1\pm\sqrt{1-C^2}).
\end{equation}

\noindent Rewriting the previous equation, we will see now a certain dualistic behaviour between the Berry phase $\gamma_{\pm}$ and the
concurrence obtained from the eigenstates $|\lambda^{\pm}(\theta,\varphi)\rangle$ of the Hamiltonian $H(\theta,\varphi)$:
the phase is always negative, either increasing or decreasing with $C$, which clear from rewriting (\ref{BerryC}) as a displaced circumference
$\big(\gamma_{\pm}/\pi+1\big)^2+C^2=1$.

\subsection{Remark on universal computational quantum gates}

The final from of $\breve{R}_{u,v}(\theta,\varphi)$ in (\ref{ybterization of Bn}) is that of a quantum gate which has been obtained
from a particular way of reaching a solution of the Yang-Baxter equation via cyclic groups. It is usually described in the literature from
topological entanglement that
a {\it universal quantum gate} is the one that entangles a product state, an entangling one.

In quantum information and computation theory, a set of universal computational quantum gates is any set of gates to which any operation possible on
a quantum computer can be reduced, that is, any other unitary operation can be expressed as a finite sequence of gates from the set. There
are examples in both one and two qubit states, some of the most famous ones being NOT or $\sqrt{NOT}$ gate for one qubit, and the CNOT
gate for two qubits. Barenco showed \cite{Barenco} that any gate can be reduced to a particular form depending on
three parameter $(\alpha,\theta,\varphi)$.  After some algebra, we obtain that any universal computational quantum gate $A(\alpha,\theta,\varphi)$ is written in
terms of spin-1/2 operators as

\begin{equation}
\frac{1}{2} \bigg( (I_{2\times2}+\sigma_z)\otimes I_{2\times2} + (I_{2\times2}-\sigma_z)\otimes e^{i[\alpha-\theta {\bf n(\varphi)} \cdot {\bf \sigma}]} \bigg),
\end{equation}

\noindent where ${\bf n(\varphi)}=(\cos\varphi,\sin\varphi,0)$ and ${\bf \sigma}=(\sigma_x,\sigma_y,\sigma_z)$.

A challenging question in the context of the present work would be how to obtain gate $A(\alpha,\theta,\varphi)$ as the evolution of
a certain physical Hamiltonian. In case we wondered if gate $A$ is a solution of the Y-B equation, direct computation in (\ref{byb}) shows that
$A$ cannot be the kind of gates obtained in such a way. This negative result implies that no hamiltonian can be obtained from any family of gates $A(\alpha,\theta,\varphi)$.

\section{A continuous family of quantum gates correspond to $\mathfrak{B}_{n, \varphi}$} \label{genral}
In this section, in the following lemma, we provide a more general family of quantum gates than the family $\mathfrak{B}_{n, \varphi}$ which was obtained from cyclic group $C_n$ in Section \ref{Bn}. The family $\mathfrak{B}_{n, \varphi}$ whose entries are parametrized by integers, $n$, is a {\it discrete} subfamily of this {\it continuous} family parametrized by real numbers, $\theta$. In fact, we will obtain $\breve{R}(\theta,\varphi)$ of \eqref{ybterization of Bn}, directly, without going through the Yang-Baxterization process of Section \ref{ybxization}. \newline

\begin{lemma} \label{generallemma}
\begin{item}
\item[(a)] Any $4$ by $4$ matrix $R_{\alpha,\beta, q}$ of the following form, where, $\alpha$, $\beta$, and $q \neq 0$ are arbitrary complex numbers, is a solution to Y-B equation \eqref{byb}
\begin{align}\label{generalR}
R_{\alpha,\beta, q}&=
\begin{pmatrix}
\alpha&0&0&\beta q \\
0&-\beta&\alpha&0 \\
0&\alpha&-\beta&0 \\
\beta q^{-1}&0&0&\alpha
\end{pmatrix}\\\nonumber
&=\alpha\begin{pmatrix}
1&0&0& 0\\
0&0&1&0 \\
0&1&0&0 \\
0&0&0&1
\end{pmatrix}+
\beta\begin{pmatrix}
0&0&0& q \\
0&-1&0&0 \\
0&0&-1&0 \\
 q^{-1}&0&0&0
\end{pmatrix}
\end{align}
\item[(b)] Let $|q|=1$. The matrix $R_{\alpha,\beta, q}$ is unitary if and only if,
\item[(i)]$\alpha \bar{\alpha} + \beta \bar{\beta}=1$
\item[(ii)]$\alpha \bar{\beta} + \beta \bar{\alpha}=0$
\end{item}

\end{lemma}


\proof Direct calculations (lengthy but still straightforward for part (a)) prove both parts of the lemma.  \newline

Conditions $(i)$ and $(ii)$ in part $(b)$ in Lemma \eqref{generallemma} leads to the following cases. \newline

If we write $\alpha=r e^{it}$ or $\beta=r' e^{it'}$, then, condition $(i)$ simply means,
\[r^2+r'^2=1 \label{i''}\]
Condition $(ii)$ implies one of the following cases: \newline

{$\bf Case \, 1:$}\newline
One of $r$ or $r'$ is zero, i.e. one of   $\alpha$ or $\beta$ is zero (not both, because that would contradict $(i)$). This in turn implies that we have: Either,
\[
R_{0,\beta, q}=
{\footnotesize
\arraycolsep=2pt
\medmuskip = 1mu
\begin{pmatrix}
0&0&0&\beta q \\
0&-\beta&0&0 \\
0&0&-\beta&0 \\
\beta q^{-1}&0&0&0
\end{pmatrix}
=\beta \begin{pmatrix}
0&0&0& q \\
0&-1&0&0 \\
0&0&-1&0 \\
 q^{-1}&0&0&0
\end{pmatrix}}; \,\, |\beta|=1 \nonumber\]
Or
\[
R_{\alpha,0, q}=
\begin{pmatrix}
\alpha&0&0&0 \\
0&0&\alpha&0 \\
0&\alpha&0&0 \\
0&0&0&\alpha
\end{pmatrix}
=\alpha\begin{pmatrix}
1&0&0&0 \\
0&0&1&0 \\
0&1&0&0 \\
0&0&0&1
\end{pmatrix}; \quad |\alpha|=1 \nonumber \]

{$\bf Case \, 2:$}\newline
If non of $r$ and $r'$ are zero then (from  $(ii)$) we must have,
\[t'=t + \frac{\pi}{2} \label{ii''}\]
Which means $\beta= r'i e^{it}$.
Therefore,
\begin{align}\label{Ralphbetq}
R_{\alpha,\beta, q}&=
e^{it}\begin{pmatrix}
r&0&0&ir' q \\
0&-ir'&r&0 \\
0&r&-ir'&0 \\
ir' q^{-1}&0&0&r
\end{pmatrix}
\end{align}
In which, with permission from relation \eqref{i''}, we can write $r=\cos \theta$ and $r'=\sin \theta$, for some real angle $\theta$. We can also write $q$ as $q=e^{i \varphi}$, for some "phase" angle $\varphi$. Thus we have,
\begin{align}\label{Rxtethaphi}
{\footnotesize
\arraycolsep=2pt
\medmuskip = 1muR_{\alpha,\beta, q}=R_{t, \theta, \varphi}=
e^{it}\begin{pmatrix}
\cos \theta &0&0&-i\sin \theta e^{i \varphi}\\
0&i\sin \theta&\cos \theta&0 \\
0&\cos \theta&i\sin \theta&0 \\
-i\sin \theta e^{-i \varphi}&0&0&\cos \theta
\end{pmatrix}}\newline
\end{align}

If we let $t=0$, and $\theta=-\frac{2\pi}{n}, \,\, n \geq 2$, in \eqref{Rxtethaphi},  we will recover $\mathfrak{B}_{n, \varphi}$ \eqref{deformedBn} obtained from cyclic group of order $n$, in Section \ref{Bn}.

Let us compare matrices in \eqref{Ralphbetq} and \eqref{Rxtethaphi} with the one in \eqref{deformedBn}, more closely. We notice that, on one hand in \eqref{deformedBn} we had $a=\cos (-\frac{2\pi}{n})$ and $-b=\sin (-\frac{2\pi}{n})$ (we consider them this way for the sake of this comparison) for integers $n\geq 2$. Whereas in \eqref{Ralphbetq} and \eqref{Rxtethaphi} we have $r=\cos \theta$ and $r'=\sin \theta$, for any real value $\theta$. On the other hand a coefficient $e^{it}$ is multiplied to the matrix. But the latter change is only scaling a matrix by $e^{it}$ which in general always creates a new unitary solution to Y-B equation from an old one. Therefore the real gain is that, we have moved from discrete angles, $\frac{2\pi}{n}$, to continues angles, $\theta$. \newline

Also notice, when $t=0$ in \eqref{Rxtethaphi}, $R_{0, \theta, \varphi}$ recaptures the matrix $\breve{R}(\theta,\varphi)$  \eqref{ybterization of Bn} in Section \ref{hamilton} directly, without going through the Yang-Baxterization process of Section \ref{ybxization}.

\begin{lemma}
The quantum gate $R_{\alpha,\beta, q}$, \eqref{generalR}, is entangling  universal gate, if and only if $\alpha$ and $\beta$ both are non-zero.
\end{lemma}

\proof
Since,
$R_{\alpha,\beta, q} (|00\rangle)
= \alpha|00\rangle +\beta |11\rangle
\nonumber$, simple calculations shows that $R_{\alpha,\beta, q}$ is entangling if and only if $\alpha$ and $\beta$ both are non-zero. Then using the Brylinskis's Theorem \cite{brylinskis 02} the proof is complete.

\section{Quantum gates correspond to cyclic groups, grading not forgotten} \label{Cnsecondapproach}
In Section \ref{Bn} when we obtain $\mathfrak{B}_n$ from representation of the quasitriangular structure on the Hopf algebra $H=\mathbb{C}C_n$, we in fact, forget the grading, in the sense explained below. In this section we show that if we don't forget the grading we can obtain different set of quantum gates from the same structure.

Let $H=\mathbb{C}C_n$ with the quasitriangular structure defined by  \cite{majidbook},
\begin{equation}
 R= \frac{1}{n}\sum_{a,b=0}^{n-1} \omega^{-ab}s^a \otimes s^b
\end{equation}
where $s$ is the generator of $C_n$, satisfying relations $s^n=s^0=1$, and $\omega=e^{\frac{2\pi i}{n}}$.

The category, $\mathcal{C}$ of all left $H$-modules, is known as the category of anyonic vector spaces \cite{majidbook}. The objects of $\mathcal{C}$ are of the form $V=\bigoplus_{i=0}^{n-1}\, V_i$. They are $C_n$-graded representations of $\mathbb{C} C_n$ and the action of $C_n$ on $V$ is given by,
\[s\rhd v= \omega^{|v|} v, \label{actofgonv} \nonumber\]
where $|v|=k$ is the degree of the homogeneous elements $v$ in $V_k$. The morphisms of $\mathcal{C}$ are linear maps that preserve the grading.\newline

The (so called) braiding map in $\mathcal{C}$ is an isomorphism $\psi_{V \otimes W}: V \otimes W \rightarrow W \otimes V$, for any two objects $V$ and $W$ in  $\mathcal{C}$, defined by,
\[\psi_{V \otimes W} (v \otimes w) = \omega^{|v||w|}\,\, w\otimes v, \label{anyonbraid} \]
where $|v|$ and $|w|$ are the degrees of homogeneous elements $v$ and $w$ in objects $V$ and $W$, respectively. We are interested in the case when $W=V$. It is well known that $\psi_{V}: V \otimes V \rightarrow V \otimes V$ provides a solution to Y-B equation \eqref{byb}. Notice that here $V$ is a $C_n$-graded vector space, and $\psi_{V}$ preserves the grading. Therefore the grading is not forgotten. \newline

Now, let $V$ be a two dimensional $C_n$-graded vector space, with the basis element $|0\rangle$ living in degree  $||0\rangle|$ and the basis element $|1\rangle$ living in degree  $||1\rangle|$. Then from formula \eqref{anyonbraid}, the $4$ by $4$ matrix representation of $\psi_{V}$, in the standard basis $\{|00\rangle, |01\rangle,|10\rangle,|11\rangle\}$  for $V \otimes V$, is:
\[ \mathfrak{B}_{\psi}=
\begin{pmatrix}
\omega^{||0\rangle|^2}&0&0& 0\\
0&0&\omega^{||0\rangle|||1\rangle|}&0 \\
0&\omega^{||0\rangle|||1\rangle|}&0&0 \\
0&0&0&\omega^{||1\rangle|^2}
\end{pmatrix}
\]
One can easily check that the gate $\mathfrak{B}_{\psi}$ is entangling, and therefore by [Berlinsky] theorem a universal gate, if and only if, $||0\rangle|^2+ ||1\rangle|^2 \neq 2||0\rangle|||1\rangle|$.

\begin{remark} We notice that, for cyclic group $C_n$, the family of quantum gates $\mathfrak{B}_{\psi}$  obtained by this approach, in a sense, less interesting than the family $\mathfrak{B}_n$ obtained in Section \ref{Bn} by forgetting the grading. However, the approach of the present section, if applied to other examples of quasitriangular Hopf algebras, might give interesting families of quantum gates. We will explore this, elsewhere. \end{remark}

\section{Conclusions}\label{Conclusions}
First, we have obtained in full detail an infinite family of quantum two qubit gates coming from the quasi-triangular Hopf algebra of the cyclic groups $C_n$, of any order $n$. A suitable $q$-deformation and the Yang-Baxterization correspond to those gates, enabled us to take two approaches for the ensuing physical Hamiltonian, responsible for the quantum entangling gate. Consequently, we have presented the concomitant physical interpretation and an application in obtaining the so-called Berry phase.

Next, we have introduced a more general family of quantum gates, which are the continuous version of the discrete family obtained from cyclic groups. 

Finally we have shown that by not forgetting the {\it grading} imposed by the action of $C_n$, we obtain a different family of quantum two qubit gates correspond to the cyclic groups $C_n$.

\section*{Acknowledgements}

J. Batle acknowledges fruitful discussions with J. Rossell\'o, Maria del Mar Batle and Regina Batle.
R. O. acknowledges support from High Impact Research MoE Grant UM.C/625/1/HIR/MoE/CHAN/04 from the
Ministry of Education Malaysia.

\section*{Appendix. Proof of Lemma \ref{Bn first approach}}

Here we shall restate the Lemma \ref{Bn first approach} and give a detailed proof of it. \newline

{$\mathbf {Lemma\,\, \ref{Bn first approach} \,\,restated:}$}
Let $C_n=\{1, s, s^2,\cdots ,s^{n-1}\}$ be the cyclic group of order $n$, with the generator $s$, satisfying the relation $s^n=s^0=1$. Let $\omega=e^{\frac{2\pi i}{n}}$, and,
\begin{equation}
 R= \frac{1}{n}\sum_{a,b=0}^{n-1} \omega^{-ab}s^a \otimes s^b
\end{equation}
be the, well known \cite{majidbook}, non-trivial quasitriangular structure on the group Hopf algebra $\mathbb{C}C_n$. If we use the following $2$ by $2$ matrix representation of elements of $C_3$,
\[s^a=\begin{pmatrix}
\cos {\frac{2a\pi}{n}}& -\sin {\frac{2a\pi}{n}}\\
\sin {\frac{2a\pi}{n}}& \cos {\frac{2a\pi}{n}}
\end{pmatrix}, \qquad a=0,1,\cdots , n-1 \]
then we will obtain a $4$ by $4$ representation of $R$ that has the following form,
\begin{align}
R=
\begin{pmatrix}
\alpha&0&0&\beta \\
0&\alpha&-\beta&0 \\
0&-\beta&\alpha&0 \\
\beta&0&0&\alpha
\end{pmatrix}
\end{align}
with, $\alpha=\cos {\frac{2\pi}{n}}$ and $\beta= -i\sin {\frac{2\pi}{n}}$. In other words,
\begin{align}
R=
\begin{pmatrix}
\cos {\frac{2\pi}{n}}&0&0&-i\sin {\frac{2\pi}{n}} \\
0&\cos {\frac{2\pi}{n}}&i\sin {\frac{2\pi}{n}}&0 \\
0&i\sin {\frac{2\pi}{n}}&\cos {\frac{2\pi}{n}}&0 \\
-i\sin {\frac{2\pi}{n}}&0&0&\cos {\frac{2\pi}{n}}
\end{pmatrix}
\end{align}

\proof
We need to have two slightly different approaches for when $n$ is even and when is odd. We proceed with odd case first. But many of the steps that we take in the proof of odd case will also be useful in the even case (especially Fact 3, below).

\subsection{$n$ odd}
When $n$ is odd, we partition the $n^2$ possible summands $\omega^{-ab}s^a \otimes s^b$ in $R$ by writing the terms correspond to $a=b=0$ and $a=0, b\neq 0$ first, followed by the rest of terms, as follows,
\begin{align}\label{rodd}
R &= \frac{1}{n}[\, 1\otimes1+\sum_{b=1}^{\frac{n-1}{2}} [1 \otimes s^b + 1\otimes s^{n-b}]\\ \nonumber
&+\sum_{a=1}^{\frac{n-1}{2}}\,\,\,\sum_{b=0}^{n-1} \omega^{-ab}[s^a \otimes s^b + s^{n-a} \otimes s^{n-b}]]
\end{align}

Next we observe the following facts, which play a crucial role in our subsequent calculations.

{$\mathbf {Fact \,1:}$}
The aim here is to show that, in terms of above representation, $s^a \otimes s^b + s^{n-a} \otimes s^{n-b}$ has the specific form  in \eqref{F1}, below. \newline

First we notice that $s^a$ which, for simplicity, for the moment, we show by $\begin{pmatrix}
A& -B\\
B&  A
\end{pmatrix}$,  corresponds to $\omega^a$. Therefore $s^{n-a}$ will correspond to $\omega^{-a}$ which is equal to the conjugate of $\omega^a$. This means, in terms of representation, $s^{n-a}=(s^a)^T =\begin{pmatrix}
A& B\\
-B&  A
\end{pmatrix}$.
Hence, for any $s^a=\begin{pmatrix}
A& -B\\
B&  A
\end{pmatrix}$ and $s^b=\begin{pmatrix}
C& -D\\
D&  C
\end{pmatrix}$, we have,
\begin{align}\label{F1}
 s^a \otimes s^b + s^{n-a} \otimes s^{n-b}=2\begin{pmatrix}
AC&0&0&BD \\
0&AC&-BD&0 \\
0&-BD&AC&0 \\
BD&0&0&AC
\end{pmatrix}
\end{align}

{$\mathbf {Fact \,2:}$}
Since, in terms of representation, $s^0=I$, it is obvious that, in \eqref{rodd}, the representation of the term $1\otimes 1$ is the $4$ by $4$ identity matrix $I$. Also the terms $1 \otimes s^b + 1\otimes s^{n-b}$, in terms of representation, are diagonal matrices. \\

So far, based on Fact 1 and Fact 2, we know that $R$, in terms of representation, has the form,
\begin{align}
R=
\begin{pmatrix}
\alpha&0&0&\beta \\
0&\alpha&-\beta&0 \\
0&-\beta&\alpha&0 \\
\beta&0&0&\alpha
\end{pmatrix}
\end{align}

To prove the specific formulas, $\alpha=\cos {\frac{2\pi}{n}}$ and $\beta=-i\sin {\frac{2\pi}{n}}$, we proceed as follows.

{$\mathbf {Fact \,3:}$}
We claim that,
\begin{align} \label{lastpartofR}
&\sum_{a=1}^{\frac{n-1}{2}}\,\,\sum_{b=0}^{n-1} \omega^{-ab}[s^a \otimes s^b + s^{n-a} \otimes s^{n-b}]\\
&=
n\begin{pmatrix} \nonumber
\cos {\frac{2\pi}{n}}&0&0&-i\sin {\frac{2\pi}{n}} \\
0&\cos {\frac{2\pi}{n}}&i\sin {\frac{2\pi}{n}}&0 \\
0&i\sin {\frac{2\pi}{n}}&\cos {\frac{2\pi}{n}}&0 \\
-i\sin {\frac{2\pi}{n}}&0&0&\cos {\frac{2\pi}{n}}
\end{pmatrix}
\end{align}

To prove this, if we put back $s^a=\begin{pmatrix}
\cos {\frac{2a\pi}{n}}& -\sin {\frac{2a\pi}{n}}\\\nonumber
\sin {\frac{2a\pi}{n}}& \cos {\frac{2a\pi}{n}}
\end{pmatrix}$ and $s^b=\begin{pmatrix}
\cos {\frac{2b\pi}{n}}& -\sin {\frac{2b\pi}{n}}\\\nonumber
\sin {\frac{2b\pi}{n}}& \cos {\frac{2b\pi}{n}}
\end{pmatrix}$ in \ref{F1}, and use the familiar trigonometric identity $\cos x \cos y=\frac{1}{2}(\cos (x-y)+\cos (x+y))$ and $\sin x \sin y=\frac{1}{2}(\cos (x-y)-\cos (x+y))$, followed by $\cos x=\frac{1}{2}(e^{ix}+e^{-ix})$ and $\sin x=\frac{1}{2i}(e^{ix}-e^{-ix})$ we get:
{\footnotesize
\arraycolsep=3pt
\medmuskip = 1mu
\begin{align}
s^a \otimes s^b + s^{n-a} \otimes s^{n-b}=\begin{pmatrix}
X+Y&0&0&X-Y \\
0&X+Y&-(X-Y)&0 \\
0&-(X-Y)&X+Y&0 \\
X-Y&0&0&X+Y
\end{pmatrix}
\end{align}}

where $X=\cos (\frac{2\pi(a-b)}{n})=\frac{1}{2}(e^{\frac{2\pi i (a-b)}{n}}+e^{-\frac{2\pi i(a-b)}{n}})$ and $Y=\cos (\frac{2\pi(a+b)}{n})=\frac{1}{2}(e^{\frac{2\pi i (a+b)}{n}}+e^{-\frac{2\pi i(a+b)}{n}})$. Note that $X$ and $Y$ depend on $a$ and $b$. \\

From this, we see that the representation of the term $\sum_{a=1}^{\frac{n-1}{2}}\,\sum_{b=0}^{n-1} \omega^{-ab}[s^a \otimes s^b + s^{n-a} \otimes s^{n-b}]$, in \eqref{rodd}, is a $4$ by $4$ matrix, in which, the main diagonal contains the element $\sum_{a=1}^{\frac{n-1}{2}}\,\sum_{b=0}^{n-1}\omega^{-ab}(X+Y)$, and the other diagonal contains the element $\sum_{a=1}^{\frac{n-1}{2}}\,\sum_{b=0}^{n-1}\omega^{-ab}(X-Y)$ repeated twice with positive signs and twice with negative sings. Let us simplify these elements.  Recall, $\omega^{-ab}=e^{\frac{2\pi i(-ab)}{n}}$.
\begin{align}\label{f1}\nonumber
&\sum_{a=1}^{\frac{n-1}{2}}\,\sum_{b=0}^{n-1}\omega^{-ab}(X+Y)\\\nonumber
&=\sum_{a=1}^{\frac{n-1}{2}}\,\sum_{b=0}^{n-1} e^{\frac{2\pi i(-ab)}{n}}[\frac{1}{2}(e^{\frac{2\pi i (a-b)}{n}}+e^{-\frac{2\pi i(a-b)}{n}})\\\nonumber
&\qquad \qquad +\frac{1}{2}(e^{\frac{2\pi i (a+b)}{n}}+e^{-\frac{2\pi i(a+b)}{n}})]\\\nonumber
&= \frac{1}{2}\sum_{a=1}^{\frac{n-1}{2}}\,\sum_{b=0}^{n-1}  [e^{\frac{2\pi i (a+b(-a-1))}{n}}+e^{-\frac{2\pi i(a+b(a-1))}{n}}\\
&\qquad \qquad +e^{\frac{2\pi i (a+b(-a+1))}{n}}+e^{-\frac{2\pi i(a+b(a+1))}{n}}]
\end{align}

Similarly,
\begin{align}\label{f2}\nonumber
&\sum_{a=1}^{\frac{n-1}{2}}\,\sum_{b=0}^{n-1}\omega^{-ab}(X-Y)\\\nonumber
&= \frac{1}{2}\sum_{a=1}^{\frac{n-1}{2}}\,\sum_{b=0}^{n-1}  [e^{\frac{2\pi i (a+b(-a-1))}{n}}+e^{-\frac{2\pi i(a+b(a-1))}{n}}\\
&\qquad \qquad -e^{\frac{2\pi i (a+b(-a+1))}{n}}-e^{-\frac{2\pi i(a+b(a+1))}{n}}]
\end{align}
Now we use the following {\it not so familiar} trigonometric identities \cite{knap 09}. Suppose $a$ and $d$ are real and $d\neq 0$, then,
\begin{align}\label{tigidentity0}\nonumber
& \sum_{b=0}^{n-1} \sin (a+bd)=\frac{\sin (\frac{nd}{2})\sin (a+\frac{(n-1)d}{2})}{\sin (\frac{d}{2})}\\
& \sum_{b=0}^{n-1} \cos (a+bd)=\frac{\sin (\frac{nd}{2})\cos (a+\frac{(n-1)d}{2})}{\sin (\frac{d}{2})}
\end{align}
Together they imply that, as long as $d\neq 0$,
\begin{eqnarray}\label{tigidentity} \nonumber
\sum_{b=0}^{n-1} e^{\pm\frac{2\pi i (a+bd)}{n}}&=&\frac{\sin (\frac{n\frac{2\pi (d)}{n}}{2})}{\sin ({\frac{\frac{2\pi(d)}{n}}{2})}} \,e^{\pm\frac{2\pi i (a+\frac{(n-1)d}{2})}{n}}\\
&=0
\end{eqnarray}
If we apply \eqref{tigidentity} into \eqref{f1} and \eqref{f2}, by taking $d$ to be $-a-1$, $a-1$, $-a+1$, $a+1$, in the four terms involved there, respectively, we will see that only the two middle terms with $d=a-1$ or $d=-a+1$, have the chance of survival only when $a=1$. Therefore \eqref{f1} and \eqref{f2} will reduce to:
\begin{eqnarray}\label{ff1}\nonumber
\sum_{a=1}^{\frac{n-1}{2}}\,\sum_{b=0}^{n-1}\omega^{-ab}(X+Y)&=& \sum_{b=0}^{n-1}  \frac{1}{2}[e^{-\frac{2\pi i}{n}}+e^{\frac{2\pi i }{n}}]\\
&=&n\cos \frac{2\pi}{n}
\end{eqnarray}

\begin{eqnarray}\label{ff2}\nonumber
\sum_{a=1}^{\frac{n-1}{2}}\,\sum_{b=0}^{n-1}\omega^{-ab}(X-Y)&=& \sum_{b=0}^{n-1}  \frac{1}{2}[e^{-\frac{2\pi i}{n}}-e^{\frac{2\pi i }{n}}]\\
&=& -n i \sin \frac{2\pi}{n}
\end{eqnarray}

Therefore we have,
\begin{align}\nonumber
&\sum_{a=1}^{\frac{n-1}{2}}\,\,\sum_{b=0}^{n-1} \omega^{-ab}[s^a \otimes s^b + s^{n-a} \otimes s^{n-b}]\\\nonumber
&=
n\begin{pmatrix} \nonumber
\cos {\frac{2\pi}{n}}&0&0&-i\sin {\frac{2\pi}{n}} \\
0&\cos {\frac{2\pi}{n}}&i\sin {\frac{2\pi}{n}}&0 \\
0&i\sin {\frac{2\pi}{n}}&\cos {\frac{2\pi}{n}}&0 \\
-i\sin {\frac{2\pi}{n}}&0&0&\cos {\frac{2\pi}{n}}
\end{pmatrix}
\end{align}
as desired.

Last thing to do is to show, the the first two parts in formula \eqref{rodd} add up to zero. This finishes the proof for $n$ odd.

By a similar (and simpler) argument as in the paragraph preceding the equation \eqref{f1}, we see that, the term $
 \sum_{b=1}^{\frac{n-1}{2}} [1 \otimes s^b + 1\otimes s^{n-b}]$ is a diagonal $4$ by $4$ matrix with the entry, $2\sum_{b=1}^{\frac{n-1}{2}} \cos(\frac{2b\pi}{n})$ repeated on the diagonal. On the other hand, again, using formula \eqref{tigidentity0}, by taking $a=0$, and $d=\frac{2\pi}{n}$ we can simplify this entry to,
\begin{align}\label{roddpart2}\nonumber
&2\sum_{b=1}^{\frac{n-1}{2}} \cos(0+b(\frac{2\pi}{n}))=2[\frac{\sin (\frac{(\frac{n+1}{2})(\frac{2\pi}{n})}{2})\cos (\frac{(\frac{n-1}{4})(\frac{2\pi}{n})}{2})}{\sin (\frac{(\frac{2\pi}{n})}{2})} \,-1]\\
&= 2[\frac{\sin (\frac{\pi}{2}+\frac{\pi}{2n})\cos(\frac{\pi}{2}-\frac{\pi}{2n})}{\sin (\frac{\pi}{n})}\, -1]=2[\frac{1}{2}-1]=-1
\end{align}
This means, in terms of representation, $\sum_{b=1}^{\frac{n-1}{2}} [1 \otimes s^b + 1\otimes s^{n-b}]$ is simply $-I$. Thus, in terms of representaton,
\begin{align}\label{roddpart12}
 1\otimes1+\sum_{b=1}^{\frac{n-1}{2}} [1 \otimes s^b + 1\otimes s^{n-b}]=I-I=0
\end{align}
Applying formulas \eqref{lastpartofR} and  \eqref{roddpart12} into \eqref{rodd}, finishes the proof of Lemma \ref{Bn first approach}, for $n$ odd.

\subsection{$n$ even}
When $n$ is even, at some point, we need to consider two cases. Either $\frac{n}{2}$ is also even (i.e. $n$ is divisible by $4$) or $\frac{n}{2}$ is odd. But before that we will make a good use of the works done in Fact 1-3.

First we partition the $n^2$ possible summands $\omega^{-ab}s^a \otimes s^b$ in $R$, by pairing them as follows.
\begin{align}\label{reven}\nonumber
R &= \frac{1}{n}[\, (1\otimes1+1\otimes s^{\frac{n}{2}})+(s^{\frac{n}{2}}\otimes 1+\omega^{-(\frac{n}{2})^2}s^{\frac{n}{2}}\otimes s^{\frac{n}{2}})\\\nonumber
&\qquad +(\sum_{b=1}^{\frac{n}{2}-1} [1 \otimes s^b + 1\otimes s^{n-b}])\\\nonumber
&\qquad +(\sum_{b=\frac{n}{2}+1}^{n-1} \omega^{-(\frac{n}{2})b}[s^\frac{n}{2} \otimes s^b + s^\frac{n}{2}\otimes s^{n-b}])\\
&\qquad +\sum_{a=1}^{\frac{n}{2}-1}\,\,\sum_{b=0}^{n-1} \omega^{-ab}[s^a \otimes s^b + s^{n-a} \otimes s^{n-b}]]
\end{align}
Based on Fact 1 and a similar argument as in Fact 2, it is clear that $R$ has the desired form. Also Fact 3 can be applied to the last (fifth) part in the formula \eqref{reven}, i.e. in terms of representation,
\begin{align}
&\sum_{a=1}^{\frac{n}{2}-1}\,\,\sum_{b=0}^{n-1} \omega^{-ab}[s^a \otimes s^b + s^{n-a} \otimes s^{n-b}]=\\\nonumber
& \qquad \qquad n\begin{pmatrix}
\cos {\frac{2\pi}{n}}&0&0&-i\sin {\frac{2\pi}{n}} \\
0&\cos {\frac{2\pi}{n}}&i\sin {\frac{2\pi}{n}}&0 \\
0&i\sin {\frac{2\pi}{n}}&\cos {\frac{2\pi}{n}}&0 \\
-i\sin {\frac{2\pi}{n}}&0&0&\cos {\frac{2\pi}{n}}
\end{pmatrix}
\end{align}
Therefore to finish the proof we need to show that the first four parts in the formula \eqref{reven} will add up to zero, in terms of representation. In the following we state some facts and use them to prove this.

{$\mathbf {Fact \,4:}$}

Since $n$ is even and $\omega^n=\omega^0=1$, we have  $\omega^{\pm\frac{n}{2}}=-1$, $\omega^{\pm\frac{n}{2}+a}=-\omega^a$ and, in terms of above presentation, $s^{\pm\frac{n}{2}+a}=-s^a$ for $a=0, 1, \cdots , n-1$. In particular for $a=0$,  and $s^{\pm\frac{n}{2}}=-I$. Also $\omega^{(a\frac{n}{2})}=\omega^{(-a\frac{n}{2})}=(-1)^a$\\

Using this fact, the first and secend parts in the formula \eqref{reven}, in terms of representation are:
\[(1\otimes1+1\otimes s^{\frac{n}{2}})=I\otimes I+I\otimes (-I)=0\]
\[(s^{\frac{n}{2}}\otimes 1+\omega^{-(\frac{n}{2})^2}s^{\frac{n}{2}}\otimes s^{\frac{n}{2}})=((-1)^{\frac{n}{2}}-1)I\otimes I=((-1)^{\frac{n}{2}}-1)I \label{revensecondpart}\]

Now we move to the third and forth parts. The forth part in \eqref{reven} can be written as follows,
\begin{align*}
&(\sum_{b=\frac{n}{2}+1}^{n-1} \omega^{-(\frac{n}{2})b}[s^\frac{n}{2} \otimes s^b + s^\frac{n}{2}\otimes s^{n-b}])\\
&=\sum_{b=\frac{n}{2}+1}^{n-1} (-1)^{b}[(-I) \otimes (-s^{b-\frac{n}{2}}) + (-I)\otimes (-s^{n-(b-\frac{n}{2})})]\\
&=\sum_{b'=1}^{\frac{n}{2}-1} (-1)^{\frac{n}{2}}(-1)^{b'}[(-I) \otimes (-s^{b'}) + (-I)\otimes (-s^{n-b'})]\\
&=\sum_{b=1}^{\frac{n}{2}-1} (-1)^{\frac{n}{2}}(-1)^{b}[I \otimes s^{b} + I\otimes s^{n-b}]
\end{align*}
Here in the third line we have made a change of variable $b'=b-\frac{n}{2}$. Now if we add this to the third part, we have, for the third and forth part in \eqref{reven},
\arraycolsep=3pt
\medmuskip = 1mu
\begin{align}\label{revenforthpart}\nonumber
&\sum_{b=1}^{\frac{n}{2}-1} [I \otimes s^b + I\otimes s^{n-b}]+\sum_{b=1}^{\frac{n}{2}-1} (-1)^{\frac{n}{2}}(-1)^{b}[I \otimes s^{b} + I\otimes s^{n-b}]\\
&=\sum_{b=1}^{\frac{n}{2}-1} (1+(-1)^{\frac{n}{2}}(-1)^{b})[I \otimes s^{b} + I\otimes s^{n-b}]=\\\nonumber
&\sum_{b=1}^{\frac{n}{2}-1}
2(1+(-1)^{\frac{n}{2}}(-1)^{b}){\footnotesize\begin{pmatrix}
\cos {\frac{2b\pi}{n}}&0&0&0 \\
0&\cos {\frac{2b\pi}{n}}&0&0 \\
0&0&\cos {\frac{2b\pi}{n}}&0 \\
0&0&0&\cos {\frac{2b\pi}{n}}
\end{pmatrix}}
\end{align}
The last step is also a result of Fact 1.
Now we are in the position to prove that the right hand sides of equations \eqref{revensecondpart} and \eqref{revenforthpart} add up to zero. Here is when we need to consider two following cases:\\

{{$\mathbf {Case 1, \frac{n}{2}\, is\, even:}$} In this case the outcome of the equation \eqref{revensecondpart} is zero, and also the outcome of the equation  \eqref{revenforthpart} is equal to,
\begin{align}\label{revenforthpart2}\nonumber
&\sum_{b=1}^{\frac{n}{2}-1}
2(1+(-1)^{b})\begin{pmatrix}
\cos {\frac{2b\pi}{n}}&0&0&0 \\
0&\cos {\frac{2b\pi}{n}}&0&0 \\
0&0&\cos {\frac{2b\pi}{n}}&0 \\
0&0&0&\cos {\frac{2b\pi}{n}}
\end{pmatrix}\\\nonumber
&=\sum_{b=2,\,\, b\,\, even}^{\frac{n}{2}}
4\begin{pmatrix}
\cos {\frac{2b\pi}{n}}&0&0&0 \\
0&\cos {\frac{2b\pi}{n}}&0&0 \\
0&0&\cos {\frac{2b\pi}{n}}&0 \\
0&0&0&\cos {\frac{2b\pi}{n}}
\end{pmatrix}=0
\end{align}
The last equality is based on the following fact, which could be verified, easily.

{{$\mathbf {Fact \,5:}$}  When $n$ and $\frac{n}{2}$ both are even,
\[\sum_{b=2,\,\, b\,\, even}^{\frac{n}{2}}\cos {\frac{2b\pi}{n}}=0\]

{{$\mathbf {Case 2, \frac{n}{2}\, is\, odd:}$}In this case the outcome of the equation \eqref{revensecondpart} is $-2I$. Also the outcome of the equation  \eqref{revenforthpart} is equal to,
\begin{align}\label{revenforthpart2}\nonumber
&\sum_{b=1}^{\frac{n}{2}-1}
2(1-(-1)^{b})\begin{pmatrix}
\cos {\frac{2b\pi}{n}}&0&0&0 \\
0&\cos {\frac{2b\pi}{n}}&0&0 \\
0&0&\cos {\frac{2b\pi}{n}}&0 \\
0&0&0&\cos {\frac{2b\pi}{n}}
\end{pmatrix}\\\nonumber
&=\sum_{b=1,\,\, b\,\, odd}^{\frac{n}{2}}
4\begin{pmatrix}
\cos {\frac{2b\pi}{n}}&0&0&0 \\
0&\cos {\frac{2b\pi}{n}}&0&0 \\
0&0&\cos {\frac{2b\pi}{n}}&0 \\
0&0&0&\cos {\frac{2b\pi}{n}}
\end{pmatrix}=2I
\end{align}
The last equality is based on the following fact, which could be verified, easily.

{{$\mathbf {Fact \,6:}$}  When $n$ is even and $\frac{n}{2}$ is odd,
\[\sum_{b=2,\,\, b\,\, odd}^{\frac{n}{2}}\cos {\frac{2b\pi}{n}}=\frac{1}{2}\]
This finishes the proof.

\end{document}